\RequirePackage{ifpdf}
\ifpdf 
\documentclass[pdftex]{sigma}
\else
\documentclass{sigma}
\fi

\def\eps{\varepsilon}
\def\ltag#1#2{\label{#1}\tag{\mbox{$#2$}}}

\begin{document}

\allowdisplaybreaks

\renewcommand{\PaperNumber}{093}

\FirstPageHeading

\renewcommand{\thefootnote}{$\star$}

\ShortArticleName{On the One Class of Hyperbolic Systems}

\ArticleName{On the One Class of Hyperbolic Systems\footnote{This
paper is a contribution to the Vadim Kuznetsov Memorial Issue
``Integrable Systems and Related Topics''. The full collection is
available at
\href{http://www.emis.de/journals/SIGMA/kuznetsov.html}{http://www.emis.de/journals/SIGMA/kuznetsov.html}}}

\Author{Vsevolod E. ADLER and Alexey B. SHABAT}

\AuthorNameForHeading{V.E. Adler and A.B. Shabat}
\Address{L.D. Landau Institute for Theoretical Physics,\\
1A prosp. ak. Semenova, 142432 Chernogolovka, Russia}

\Email{\href{mailto:adler@itp.ac.ru}{adler@itp.ac.ru},
\href{mailto:shabat@itp.ac.ru}{shabat@itp.ac.ru}}

\ArticleDates{Received October 27, 2006; Published online December 27, 2006}

\Abstract{The classif\/ication problem is solved for some type of
nonlinear lattices. These lattices are closely related to the
lattices of Ruijsenaars--Toda type and def\/ine the B\"acklund
auto-transformations for the class of two-component hyperbolic
systems.}


\Classification{35L75; 35Q55; 37K10; 37K35}

\begin{flushright}
\it This paper is dedicated to the memory of Vadim Kuznetsov
\end{flushright}

\section{Introduction}\label{sec:intro}

In this paper we solve the problem of classif\/ication of the
consistent pairs of the lattices
\begin{gather}
\label{uvx} u_x=F(u_1,u,v),\qquad  v_x=G(u,v,v_{-1}),
\\[.5ex]
\label{uvy} u_y=P(u_{-1},u,v),\qquad  v_y=Q(u,v,v_1).
\end{gather}
Here $u=u(n,x,y)$, $v=v(n,x,y)$, $n\in\mathbb Z$ and subscripts
denote both the partial derivatives with respect to $x,y$ and the
shift with respect to $n$. The nondegeneracy conditions are
assumed
\begin{gather}\label{ne0}
 F_vF_{u_1}G_uG_{v_{-1}}P_vP_{u_{-1}}Q_uQ_{v_1}\ne0.
\end{gather}
Due to the condition $F_v\ne0$, the f\/irst equation of the
lattice (\ref{uvx}) can be solved with respect to the variable
$v$, then the second equation of the lattice rewrites as a lattice
of the Ruijsenaars--Toda type \cite{Ruijsenaars1986,
Ruijsenaars1990, BruschiRagnisco1989, Suris1990}
\begin{gather}\label{RTLxx}
 u_{xx}=A(u_{1,x},u_x,u_{-1,x},u_1,u,u_{-1})
\end{gather}
while its symmetry takes the form
\begin{gather}\label{RTLxy}
 u_y=B(u_x,u_1,u,u_{-1}).
\end{gather}
Clearly, the roles of $x$ and $y$ may be interchanged and equation
(\ref{RTLxx}) may be replaced by a lattice of the form
\begin{gather}\label{RTLyy}
 u_{yy}=C(u_{1,y},u_y,u_{-1,y},u_1,u,u_{-1}).
\end{gather}
The classif\/ication problem for the lattices (\ref{RTLxx}) was
solved, in a dif\/ferent setting, in our
paper~\cite{AdlerShabat1997a}, see also \cite{AdlerShabat1997b,
AdlerShabat1998, MarikhinShabat1999, AdlerMarikhinShabat2001}. The
approach based on the relation to the lattices (\ref{uvx}) allows
to reproduce this result in a slightly more general way and is
promising for further generalizations. The classif\/ication of the
integrable lattices of the form (\ref{uvx}) was obtained by
Yamilov \cite{Yamilov2000}. The main postulate in his work was
that, instead of (\ref{uvy}), a symmetry of a high enough order
exists, and moreover, the lattice was assumed to be Hamiltonian.
Our aim here is to obtain the answer under the minimal
restrictions. However, in comparison with the result of Yamilov,
our list contains only two more pairs. These pairs are not
Hamiltonian, and they are reducible in some def\/inite sense. We
discuss these curious examples separately in Section
\ref{sec:eps}.

The interest to the lattices (\ref{uvx}) is explained by their
close relation to many other integrable models, see
e.g.~\cite{RagniscoSantini1990, ShabatYamilov1991,
AdlerYamilov1994}. The higher symmetries of such lattices generate
the hierarchies of the evolution systems of the nonlinear
Schr\"odinger type. In this context equation (\ref{uvy}) def\/ines
the f\/irst negative f\/low of the hierarchy and corresponds to
some hyperbolic system, see Section~\ref{sec:hyps}. Although not
all integrable two-component hyperbolic systems can be obtained in
this way, this correspondence is a source of important examples.
Also, the linear combinations of
the~f\/lows~(\ref{uvx}),~(\ref{uvy}) constitute a class of
equations containing Ablowitz--Ladik and Sklyanin
lat\-tices~\cite{AdlerShabatYamilov2000,Adler2000a}.

In Section \ref{sec:DTL} we demonstrate that the transform to the
Ruijsenaars--Toda lattice (\ref{RTLxx}) exhibits a hidden discrete
symmetry of the lattice (\ref{uvx}). Namely, it turns out that the
lattices of the form~(\ref{RTLxx}) for the variables $u$ and $v$
coincide (after an appropriate substitution), and this allows to
expand the equations onto the square grid and brings to the
discrete Toda lattices \cite{Hirota, Suris95, Adler2000b}. It
should be noted that conversely, the lattices of the form
(\ref{uvx}) or (\ref{RTLxx}) can be obtained from the discrete
Toda type equations under the continuous limits
\cite{AdlerSuris2004}.

Section \ref{sec:lists} contains the list of the consistent
Hamiltonian pairs (\ref{uvx}), (\ref{uvy}) and the corresponding
lists of the hyperbolic systems and Ruijsenaars--Toda lattices.

Returning to the setting of the problem, notice that it is natural
to consider equivalent lattices related by the point substitutions
\begin{gather}\label{newuv}
 \tilde u=\phi(u),\qquad \tilde v=\psi(v)
\end{gather}
the scalings $\tilde x=\alpha x$, $\tilde y=\beta y$ and the
renamings
\begin{gather}
\label{sym_uv}
 u\leftrightarrow v,\qquad n\leftrightarrow -n,\qquad
 F\leftrightarrow G,\qquad P\leftrightarrow Q,
 \\
\label{sym_xy}
 x\leftrightarrow y,\qquad n\leftrightarrow -n,\qquad
 F\leftrightarrow P,\qquad G\leftrightarrow Q.
\end{gather}
We use these transforms in order to bring a lattice to a simpler
form. Moreover, the substitution
\begin{gather}\label{shift}
 \tilde u=u+\alpha x+\beta y+\gamma n,\qquad \tilde v=v+\alpha x+\beta y+\gamma n
\end{gather}
is often useful. It is clear that it brings, in general, to a
nonautonomous lattice. However, if the right hand sides of the
lattices contain only the dif\/ferences $u-v$, $u_{\pm1}-v$ then
this transformation preserves the class under consideration. For
brevity, we refer to all mentioned transformations as to
admissible substitutions. Our main result is the following
theorem.

\begin{theorem}\label{th:list}
The consistent lattices \eqref{uvx}, \eqref{uvy}, such that the
nondegeneracy conditions \eqref{ne0} are fulfilled, are brought by
the admissible substitutions to one of the Hamiltonian pairs in
the list~{\rm \ref{list}},
\begin{gather}
\label{H}
 u_x=a(u,v)\delta_vH,\qquad v_x=-a(u,v)\delta_uH,\qquad H=K(u_{n+1},v_n)+L(u_n,v_n),
 \\
\label{R}
 u_y=a(u,v)\delta_vR,\qquad v_x=-a(u,v)\delta_uR,\qquad R=M(u_n,v_{n+1})+N(u_n,v_n)
\end{gather}
($\delta_uH=\partial_u\sum_nT^n(H)$ denotes the lattice
variational derivative),  or to one of the pairs
\begin{gather}
\ltag{X7}{X_7} u_x=\frac{v+(1+\eps)u+u_1}{\eps v-u_1}, \qquad
v_x=\frac{v_{-1}+(1+\eps^{-1})v+u}{v_{-1}-\eps^{-1}u},
\\
\ltag{Y7}{Y_7}
u_y=\frac{v+(1+\eps^{-1})u+u_{-1}}{\eps^{-1}v-u_{-1}}, \qquad
 v_y=\frac{v_1+(1+\eps)v+u}{v_1-\eps u},
\\
\ltag{X8}{X_8} u_x=f(v-u_1),\qquad v_x=\eps f(v_{-1}-u),
\\
\ltag{Y8}{Y_8} u_y=p(\eps u_{-1}-v),\qquad v_y=p(\eps u-v_1)
\end{gather}
or to the linear lattices.
\end{theorem}

\section{Necessary conditions}\label{sec:necc}

In this Section we deduce some consequences from the compatibility
condition of the lattices under consideration. Let $T$ denote the
shift operator $n\to n+1$.

\begin{proposition}\label{pr:FGPQ}
If the lattices \eqref{uvx}, \eqref{uvy} commute then a constant
$\eps$ and functions $a(u,v)$, $b(u,u_1)$, $c(v,v_1)$, $h(u_1,v)$,
$r(u,v_1)$ exist and are unique up to the scaling $(a,b,c,h,r)\to{}$\linebreak
$(ka,kb,kc,h/k,r/k)$, such that
\begin{gather}
\label{F} F_{u_1} =ah,  \qquad F_v =bh,
\\
\label{G} G_{v_{-1}} =-\eps aT^{-1}(h),\qquad   G_u =-T^{-1}(ch),
\\
\label{P} P_{u_{-1}} =-\eps aT^{-1}(r),\qquad   P_v =-T^{-1}(br),
\\
\label{Q} Q_{v_1} =ar,\qquad   Q_u =cr.
\end{gather}
\end{proposition}

\begin{proof}
Computing $u_{xy}$ in two ways gives the equation
\begin{gather}\label{uxy}
 F_{u_1}T(P)+F_uP+F_vQ=P_{u_{-1}}T^{-1}(F)+P_uF+P_vG.
\end{gather}
Dif\/ferentiating of this expression with respect to $u_1$ and
$u_{-1}$ yields $F_{uu_1}P_{u_{-1}}=P_{uu_{-1}}F_{u_1}$, which
implies
\begin{gather*}
 F_{u_1}=a(u,v)h(u_1,v),\qquad P_{u_{-1}}=a(u,v)\hat r(u_{-1},v).
\end{gather*}
Using the symmetry (\ref{sym_uv}) gives analogously
\begin{gather*}
 G_{v_{-1}}=\tilde a(u,v)\hat h(u,v_{-1}),\qquad Q_{v_1}=\tilde a(u,v)r(u,v_1).
\end{gather*}
Let us prove that, moreover, the following relations hold
\begin{gather}\label{bc}
 T(P_v) =-br, \qquad \tilde aF_v =abh,
 \\
 T^{-1}(Q_u) =-\hat c\hat r, \qquad aG_u  =\tilde a\hat c\hat h,
 \\
 T^{-1}(F_v) =-\hat b\hat h, \qquad \tilde aP_v  =a\hat b\hat r,
 \\
 T(G_u) =-ch, \qquad aQ_u =\tilde acr
\end{gather}
with some functions
\[
 b=b(u,u_1),\qquad \hat c=\hat c(v_{-1},v),\qquad
 \hat b=\hat b(u_{-1},u),\qquad c=c(v,v_1).
\]
In order to obtain the f\/irst line, dif\/ferentiate (\ref{uxy})
with respect to $v_1$. This yields $F_vQ_{v_1}+F_{u_1}T(P_v)=0$,
or $\tilde aF_v/(ah)=-T(P_v)/r$. Here the left and right hand
sides depend respectively on $u_1$, $u$, $v$ and $u$, $u_1$,
$v_1$, so that their common value is some function $b(u,u_1)$. In
order to prove the rest, it is suf\/f\/icient to use the
symmetries (\ref{sym_uv}), (\ref{sym_xy}).

Now, one easily obtains from (\ref{bc}) the relation
\[
 \frac{a(u,v)}{\tilde a(u,v)}\cdot\frac{b(u,u_1)}{T(\hat b(u_{-1},u))}=
 \frac{T(a(u,v))}{T(\tilde a(u,v))}\cdot\frac{c(v,v_1)}{T(\hat c(v_{-1},v))}.
\]
This implies $a/\tilde a=\psi(v)/\phi(u)$. One may assume
$a=\tilde a$ without loss of generality, taking into account the
substitution
\[
 a\to a\psi,\qquad \tilde a\to\tilde a\phi,\qquad
 h\to h/\psi,\qquad \hat r\to\hat r/\psi,\qquad
 \hat h\to\hat h/\phi,\qquad r\to r/\phi.
\]
Then $b=\eps T(\hat b)$, $c=\eps T(\hat c)$ with some constant
$\eps$ and functions $h$, $\hat h$ and $r$, $\hat r$ are related
by equations
\[
 \hat h=-\eps T^{-1}(h),\qquad \hat r=-\eps T^{-1}(r).
\]
Taking all together proves the required equations.
\end{proof}

The system of equations \eqref{F}--\eqref{Q} is overdetermined and
further information about the form of the lattices will be
obtained mainly by analysis of the compatibility conditions of
this system. However, this cannot solve the problem completely,
since the functions $F$, $P$ and~$G$,~$Q$ are found from this
system at least up to the addition of arbitrary functions on $u$
and $v$ respectively. Therefore, after some point we will need
additional information. It turns out that use of the following
proposition allows to determine the right hand sides of the
lattices up to few arbitrary constants. The f\/inal answer is
obtained then by the intermediate check of the consistency.

\begin{proposition}\label{pr:CL}
If the lattices \eqref{uvx}, \eqref{uvy} commute then the
following equations hold
\begin{gather}
\label{ahPQ}
 h_vQ+h_{u_1}T(P) =-h(T(P_u)+Q_v+k_1), \qquad a_uP+a_vQ =a(P_u+Q_v+k_1),
 \\[.5ex]
\label{arFG}
 r_uF+r_{v_1}T(G) =-r(F_u+T(G_v)+k_2), \qquad a_uF+a_vG =a(F_u+G_v+k_2),
\end{gather}
where $k_1$, $k_2$ are some constants and the functions $a$, $r$, $h$
are defined in the Proposition~{\rm \ref{pr:FGPQ}}.
\end{proposition}
\begin{proof}
Equations (\ref{ahPQ}), (\ref{arFG}) are equivalent to the
conservation laws
\begin{gather}
  D_y(\log F_{u_1})=(1-T)(P_u), \qquad
  D_y(\log G_{v_{-1}})=(1-T^{-1})(Q_v),
  \\[.5ex]
  D_x(\log P_{u_{-1}})=(1-T^{-1})(F_u),  \qquad
  D_y(\log Q_{v_1})=(1-T)(G_v),
\end{gather}
where the left hand sides are replaced in accordance to the
formulae (\ref{F})--(\ref{Q}). In its turn, the f\/irst of these
conservation laws is obtained by dif\/ferentiating of (\ref{uxy})
with respect to $u_1$, and in order to obtain the other ones it is
suf\/f\/icient to apply the symmetries (\ref{sym_uv}),
(\ref{sym_xy}).
\end{proof}

\begin{remark}
We see, comparing the equations (\ref{H}) and (\ref{F}), (\ref{G})
that for the Hamiltonian lattices $h=H_{u_1v}$ and $\eps=1$. In
the general case parameter $\eps$ remains undetermined almost till
the end of calculations. This makes necessary to consider several
additional branches with $\eps\ne1$, leading to essential
complication of the analysis (see Propositions \ref{pr:a21},
\ref{pr:a0} below). Only in two of these cases the solution is not
empty. Moreover, it is not dif\/f\/icult to prove that in the
Hamiltonian case the second column of the equations~(\ref{ahPQ}),
(\ref{arFG}) turns into identities and the constants $k_1$, $k_2$
vanish.
\end{remark}

In what follows we use the substitutions (\ref{newuv}). They act
on the functions under consideration in accordance  to the rules
\begin{gather*}
 \tilde F=\phi'(u)F,\qquad \tilde G=\psi'(v)G,\qquad
 \tilde P=\phi'(u)P,\qquad \tilde Q=\psi'(v)Q,
 \\
 \tilde h=h/(\psi'(v)\phi'(u_1)),\qquad \tilde r=r/(\phi'(u)\psi'(v_1)),
 \\
 \tilde a=\phi'(u)\psi'(v)a,\qquad \tilde b=\phi'(u)\phi'(u_1)b,\qquad
 \tilde c=\psi'(v)\psi'(v_1)c.
\end{gather*}

\section{Proof of the classif\/ication theorem}\label{sec:proof}

We will see soon that it is convenient to divide all lattices into
two classes, depending on whether $a$ is of the form
$a(u,v)=a_1(u)a_2(v)$ or not. We start from the case when this
factorization does not hold. The form of the lattices from this
subclass (which contain the larger part of the list) is def\/ined
more exactly in the following proposition.

\begin{proposition}\label{pr:uvx}
If the lattice \eqref{uvx} satisfies the equations \eqref{F},
\eqref{G} at $aa_{uv}\ne a_ua_v$ then it can be brought, by a
substitution \eqref{newuv}, to one of the following types:
\begin{gather}
\label{a1}
 \left\{\begin{aligned}
  u_x&=au_1+f_1(u)v+f(u),
  \\
  v_x&=-\eps av_{-1}+g_1(v)u+g(v),
 \end{aligned}\right.
 \qquad a=\sum^1_{i,j=0}a_{ij}u^iv^j,
 \\
\label{a2}
 \left\{\begin{aligned}
  u_x&=\frac{a}{v-u_1}-\frac{a_v}{2}+f(u),\\
  v_x&=\frac{\eps a}{v_{-1}-u}+\frac{\eps a_u}{2}+g(v),
 \end{aligned}\right.
 \qquad  a=\sum^2_{i,j=0}a_{ij}u^iv^j.
\end{gather}
\end{proposition}
\begin{proof}
Cross-dif\/ferentiating of the relations (\ref{F}) yields
\begin{gather}\label{abh}
 a_vh+ah_v=b_{u_1}h+bh_{u_1}.
\end{gather}
Dividing by $h$ and applying the operator
$\partial_v\partial_{u_1}-(\log h)_{vu_1}$ yields
\begin{gather*}
 aN_{u_1}=bM_v,\qquad
 N=\frac{h_{vv}}{h}-\frac{3h^2_v}{2h^2},\qquad
 M=\frac{h_{u_1u_1}}{h}-\frac{3h^2_{u_1}}{2h^2}.
\end{gather*}
Dif\/ferentiating with respect to $u$ gives $a_uN_{u_1}=b_uM_v$.
Since $aa_{uv}\ne a_ua_v$, hence $a_ub\ne ab_u$ and therefore
$M_v=N_{u_1}=0$, that is $M=M(u_1)$, $N=N(v)$. It is easy to
obtain the rule of transformation of these functions under the
substitution (\ref{newuv}):
\begin{gather*}
 M=(\phi')^2\tilde M+\frac{\phi'''}{\phi'}-\frac{3(\phi'')^2}{2(\phi')^2},\qquad
 N=(\psi')^2\tilde N+\frac{\psi'''}{\psi'}-\frac{3(\psi'')^2}{2(\psi')^2}.
\end{gather*}
We set $M=N=0$ by an appropriate choice of $\phi$, $\psi$. Denote
$h=H^{-2}$, then these equalities and equation (\ref{abh}) take
the form
\begin{gather}\label{abH}
 H_{vv}=0,\qquad H_{u_1u_1}=0,\qquad a_vH-2aH_v=b_{u_1}H-2bH_{u_1}.
\end{gather}
Notice that we may still apply the M\"obius transformations of $v$
and $u_1$. This allows to bring~$H$ to one of the forms $H=1$ or
$H=v-u_1$ (depending on whether the original polynomial is
reducible or not).

If $h=1$ then the system (\ref{F}) takes the form
$F_{u_1}=a(u,v)$, $F_v=b(u,u_1)$, whence
\begin{gather*}
 F=\alpha(u)vu_1+\beta(u)v+\gamma(u)u_1+\delta(u).
\end{gather*}
Analogously,
\begin{gather*}
 G=\varkappa(v)v_{-1}u+\lambda(v)v_{-1}+\mu(v)u+\nu(v),
\end{gather*}
moreover $-\eps a=G_{v_{-1}}=-\eps F_{u_1}$, and equations
(\ref{a1}) follow.

If $h=(v-u_1)^{-2}$ then one easily f\/inds from equations
(\ref{abH}) that $a=a_2v^2+a_1v+a_0$, $b=-a_2u_1^2-a_1u_1-a_0$,
where the coef\/f\/icients depend on $u$. Solving the system
(\ref{F}) yields
\begin{gather*}
 F=\frac{\alpha(u)vu_1+\beta(u)v+\gamma(u)u_1+\delta(u)}{v-u_1},
\end{gather*}
where $\alpha=a_2$, $\beta+\gamma=a_1$, $\delta=a_0$. Analogously,
\begin{gather*}
 G=\frac{\varkappa(v)v_{-1}u+\lambda(v)v_{-1}+\mu(v)u+\nu(v)}{v_{-1}-u}
\end{gather*}
and one proves that $a$ is a biquadratic polynomial by comparing
the formulae
\begin{gather*}
 a=\alpha(u)v^2+(\beta(u)+\gamma(u))v+\delta(u),\qquad
 \eps a=\varkappa(v)u^2+(\lambda(v)+\mu(v))u+\nu(v).
\end{gather*}
The lattice takes the form (\ref{a2}) with $2f=\beta-\gamma$,
$2g=\lambda-\mu$.
\end{proof}

The similar statement is valid for the lattice (\ref{uvy}) as
well, due to the symmetry (\ref{sym_xy}). However, the
substitutions (\ref{newuv}) for both lattices may be dif\/ferent,
so that the second one should be written for some another
variables $U=U(u)$, $V=V(v)$:
\begin{gather}
\label{A1} \left\{\begin{aligned}
  U_y&=-\eps AU_{-1}+p_1(U)V+p(U),
  \\
  V_y&=AV_1+q_1(V)U+q(V),
 \end{aligned}\right.
 \qquad A=\sum^1_{i,j=0}A_{ij}U^iV^j,
 \\
\label{A2} \left\{\begin{aligned}
  U_y&=\frac{\eps A}{U_{-1}-V}+\frac{\eps A_V}{2}+p(U),
  \\
  V_y&=\frac{A}{U-V_1}-\frac{A_U}{2}+q(V),
 \end{aligned}\right.
\qquad A=\sum^2_{i,j=0}A_{ij}U^iV^j.
\end{gather}
The lattices (\ref{a1}), (\ref{a2}) and (\ref{A1}), (\ref{A2})
combine into three variants of the pairs, up to the
symmet\-ry~(\ref{sym_xy}). In each case some additional
information about the right hand sides can be obtained by
comparing the formulae for the functions $a$, $b$, $c$ corresponding to
both lattices. The following lemma is useful as well. Here we
denote
\begin{gather*}
 \tilde F(U_1,U,V)=U'F(u_1,u,v),\qquad \tilde G(U,V,V_{-1})=V'G(u,v,v_{-1})
\end{gather*}
for the right hand sides of the lattice (\ref{uvx}) rewritten in
the variables $U,V$.

\begin{lemma}\label{l:linq}
If the lattice \eqref{uvx} is of the form \eqref{a1} then its
symmetry \eqref{uvy} satisfies the conditions $P_{uu}=Q_{vv}=0$.
If the lattice \eqref{uvx} is of the form \eqref{a2} then
$P_{uuu}=Q_{vvv}=0$ and, moreover, $P$~and~$Q$ coincide as the
functions on their arguments, that is $P(u,v,v_1)=Q(u,v,v_1)$.

Analogously, if the lattice \eqref{uvy} is of the form \eqref{A1}
then $\tilde F_{UU}=\tilde G_{VV}=0$, and if the lattice
\eqref{uvy} is  of the form \eqref{A2} then $\tilde F_{UUU}=\tilde
G_{VVV}=0$ and $\tilde F(U_1,U,V)=\tilde G(U_1,U,V)$.
\end{lemma}

\begin{proof}
Let us use the f\/irst of the relations (\ref{ahPQ}). If $h=1$
then the equalities $P_{uu}=Q_{vv}=0$ follow immediately. If
$h=(v-u_1)^{-2}$ then dif\/ferentiating with respect to $u_1$
gives
\begin{gather*}
 Q=T(P)+(v-u_1)T(P_u)+\frac12(v-u_1)^2T(P_{uu}).
\end{gather*}
Dif\/ferentiating once more yields $P_{uuu}=0$. Moreover, the last
equation is nothing but the Taylor expansion in the second
argument for the function $P(u,v,v_1)$. The equations for $F,G$
follow in virtue of the symmetry (\ref{sym_xy}).
\end{proof}

\begin{proposition}\label{pr:a21}
The consistent lattices with the condition $aa_{uv}\ne a_ua_v$ are
brought by the admissible substitutions to one of the pairs
$(\ref{X1},\ref{Y1})$--$(\ref{X4},\ref{Y4})$ or
$(\ref{X7},\ref{Y7})$.
\end{proposition}
\begin{proof}
1) Let the lattices be of the form (\ref{a2}), (\ref{A2}). Then
one can set $h=(v-u_1)^{-2}$ and $r=(U-V_1)^{-2}U'V'_1$, after
scaling of $y$. The comparison of the functions $a,b,c$ gives
\begin{gather}
a=a(u,v)= \frac{A(U,V)}{U'V'},\quad
 -b= a(u,u_1)= \eps \frac{A(U_1,U)}{U'_1U'},\quad
 -c= \eps a(v,v_1)= \frac{A(V_1,V)}{V'_1V'},\label{a2A2}
\end{gather}
where
\begin{gather*}
 a(u,v)=\sum^2_{i,j=0}a_{ij}u^iv^j,\qquad A(U,V)=\sum^2_{i,j=0}A_{ij}U^iV^j.
\end{gather*}
The dif\/ferentiation of the f\/irst equation yields
\begin{gather*}
 (a_uU'+aU'')V'=A_UU',\qquad
 (a_{uu}U'+2a_uU''+aU''')V'=A_{UU}(U')^2+A_UU'',
 \\[.5ex]
 (3a_{uu}U''+3a_uU'''+aU^{IV})V'=3A_{UU}U'U''+A_UU'''
\end{gather*}
and elimination of $A_U$, $A_{UU}$ brings to the relation
\begin{gather*}
 (2U'''U'-3(U'')^2)U'a_u=(4U'''U''U'-U^{IV}(U')^2-3(U'')^3)a.
\end{gather*}
Since $aa_{uv}\ne a_ua_v$ this implies $2U'''U'-3(U'')^2=0$, that
is $U$ is a linear-fractional function. The same is true for $V$
as well.

Since the lattice (\ref{A2}) is form invariant under the
substitution $\tilde U=M(U)$, $\tilde V=M(V)$ where $M$ is an
arbitrary M\"obius transform, hence one may set $U=u$ without loss
of generality. Then one may apply the M\"obius transformations to
the variables: $\tilde u=M(u)$, $\tilde v=M(v)$. Under this
transformation the function $V$ is changed in accordance to the
formula $\tilde V=MVM^{-1}$, so that it can be brought to one of
the forms $V(v)=\delta v$ or $V(v)=v+2\delta$. The system
(\ref{a2A2}) becomes equivalent to the equation $a(u,v)=\eps
A(v,u)$ which def\/ines $A$ and the relation
\begin{gather*}
 \eps a(u,v)V'(v)=a(V(v),u)
\end{gather*}
which means that function $a$ possesses some generalized symmetry
property. All biquadratic polynomials which satisfy this identity
can be found directly. They are listed in the Table \ref{a_table}
(up to the inversion $(u,v)\to(1/u,1/v)$; the restrictions on the
parameters are introduced in order to avoid the intersections and
to provide the condition $aa_{uv}\ne a_ua_v$). Now we only have to
f\/ind the coef\/f\/icients of $f$, $g$, $p$, $q$ which are
quadratic polynomials, in accordance to the Lemma \ref{l:linq}.
This is done by the direct computations which show that the most
cases are empty. As a result, we obtain the pairs
(\ref{X1},\ref{Y1}), (\ref{X7},\ref{Y7}) and also the pairs
\begin{gather}
u_x=\frac{a}{v-u_1}+u-v+\beta,\qquad \nonumber
v_x=\frac{a}{v_{-1}-u}+u-v+\beta,
\\
u_y=\frac{a}{u_{-1}-v-2\delta}-u+v+\gamma,\qquad
v_y=\frac{a}{u-v_1-2\delta}-u+v+\gamma,\label{a2.2}
\end{gather}
where $a=(u-v-\delta)^2+\alpha$, and
\begin{gather}
u_x=\frac{a}{v-u_1}-\delta v+\beta u,\qquad
v_x=\frac{a}{v_{-1}-u}+u+(\beta+\alpha)v, \nonumber
\\
u_y=\frac{\delta a}{u_{-1}-\delta v}+\delta
v+(\gamma+\alpha)u,\qquad v_y=\frac{a}{u-\delta v_1}-u+\gamma
v,\label{a2.3}
\end{gather}
where $a=u^2+\alpha uv+\delta v^2$. However, these lattices can be
brought to the particular cases of the lattice (\ref{X1},\ref{Y1})
by the transformation (\ref{shift}). For the lattices (\ref{a2.2})
one should apply the substitution
\begin{gather*}
 u\to u+(\beta+\delta)x+(\gamma-\delta)y-\delta(n-1),\qquad
 v\to v+(\beta+\delta)x+(\gamma-\delta)y-\delta n
\end{gather*}
and the new variables satisfy the lattices (\ref{X1},\ref{Y1})
with $a=(u-v)^2+\alpha$. For (\ref{a2.3}), the analogous
substitution is conjugated by exponentiation.

\begin{table}
\caption{Solutions of the equation $\eps a(u,v)V'(v)=a(V(v),u)$.}
\label{a_table}
$$\begin{array}{|r|c|l|c|}
 \hline &&&\\[-3mm]
 \eps~~ & V(v) & \qquad a(u,v) &\\
 \hline &&&\\[-2mm]
      1 &  v   & a_{22}u^2v^2+a_{12}uv(u+v)+a_{02}(u^2+v^2) & \\
        &      & \qquad +a_{11}uv+a_{01}(u+v)+a_{00} & (\ref{X1},\ref{Y1}) \\
     -1 &  v   & (u-v)(\alpha uv+\beta(u+v)+\gamma) & -\\[1mm]
 \hline &&&\\[-2mm]
      1 & v+2\delta & (u-v-\delta)^2+\alpha,\qquad \delta\ne0 & (\ref{a2.2}) \\
     -1 & v+2\delta & u-v-\delta,\qquad \delta\ne0 & - \\[1mm]
 \hline &&&\\[-2mm]
  \sqrt[3]{1} & \eps v  & \alpha u^2v^2 +\eps u+v,\qquad \alpha\ne0 & -\\
  \sqrt{-1}   & -v      & \alpha uv(u+\eps v)+\beta(\eps u +v),\qquad
                          \alpha\beta\ne0 & - \\
     -1 & -v   & \alpha u^2v^2+\beta(u^2+v^2)+\gamma,\qquad \beta\gamma\ne\alpha^2 & -\\
      1 & \delta v & u^2+\alpha uv+\delta v^2,\qquad \delta\ne0,1 & (\ref{a2.3}) \\
     -1 & \delta v & u^2-\delta v^2,\qquad \delta\ne0,1 & -\\
   &\eps^{-2}v & \eps u+v,\qquad \eps^2\ne1 & (\ref{X7},\ref{Y7}) \\[1mm]
\hline\end{array}$$
\end{table}

2) Let the lattices be of the form (\ref{a2}), (\ref{A1}),
$h=(v-u_1)^{-2}$, $r=U'V'_1$. Then
\begin{gather*}
 a(u,v)= \sum^2_{i,j=0}a_{ij}u^iv^j=\frac{A_{11}UV+A_{10}U+A_{01}V+A_{00}}{U'V'},
 \\
 b(u,u_1)= -a(u,u_1)=\frac{\eps A_{11}UU_1+\eps A_{01}U-p_1(U_1)}{U'U'_1},
 \\
 c(v,v_1)= -\eps a(v,v_1)=\frac{A_{11}VV_1+A_{10}V_1+q_1(V)}{V'V'_1}.
\end{gather*}
As in the previous case, the f\/irst equation implies that
functions $U,V$ are linear-fractional. The other two equations are
equivalent to relations
\begin{gather}
\label{aA2.1}
 \frac{A_{11}V(v)+A_{10}}{V'(v)}=-\eps\frac{A_{11}U(v)+A_{01}}{U'(v)},
 \\
\label{aA2.2}
 \frac{p_1(U(u))}{U'(u)}=\frac{A_{01}V(u)+A_{00}}{V'(u)},\qquad
 \frac{q_1(V(v))}{V'(v)}=-\eps\frac{A_{10}U(v)+A_{00}}{U'(v)}.
\end{gather}
We assume, without loss of generality, that either $A=UV+1$ or
$A=\eps U-V$. Equation (\ref{aA2.1}) implies $U(u)=cV(u)^{-\eps}$
(and, consequently, $\eps=\pm1$) in the f\/irst case, or
$U(u)=V(u)+c$ in the second one. The appropriate substitutions,
the M\"obius one in the lattice (\ref{a2}) and the linear one in
(\ref{A1}), reduce the problem to the following cases:
\begin{gather*}
\eps =1,   U=u,\qquad    V=v^{-1},\qquad a=-v(u+\delta v),\qquad
A=UV+\delta,\qquad \delta\ne0;
\\
\eps =-1,\qquad    U=u, \qquad   V=v, \qquad   a=A=uv+1;
\\
U=u, \qquad   V=v,  \qquad  a=A=\eps u-v+\delta.
\end{gather*}
Moreover, the functions $p_1,q_1$ are def\/ined from the relations
(\ref{aA2.2}) and functions $f(u)$, $g(v)$, $p(u)$, $q(V)$ from the
equations
\begin{gather*}
 f''=0,\qquad ((V'g)'/V')'=0,\qquad p'''=0,\qquad (q/V')_{vv}=0,
\end{gather*}
in accordance to the Lemma \ref{l:linq}. In the f\/irst case,
after def\/ining the coef\/f\/icients by the direct computation
and the substitutions $u\to e^u$, $v\to e^v$ and (\ref{shift}) we
obtain the lattices (\ref{X3},\ref{Y3}). The second case turns out
to be empty and the third one contains, at $\eps=1$, the pair
(\ref{X2},\ref{Y2}), up to the admissible transformations.

3) Let the lattices be of the form (\ref{a1}), (\ref{A1}), $h=1$
and $r=U'V'_1$. The comparison of the expressions for the function
$a$ brings to the equation
\begin{gather*}
(a_{11}uv+a_{10}u+a_{01}v+a_{00})U'V'=A_{11}UV+A_{10}U+A_{01}V+A_{00}.
\end{gather*}
Computing $aa_{uv}-a_ua_v$ one obtains the relation
\begin{gather*}
(a_{11}a_{00}-a_{10}a_{01})U'V'=A_{11}A_{00}-A_{10}A_{01}
\end{gather*}
which implies $U''=V''=0$. Therefore, we may assume, after the
linear substitutions, $U=u$, $V=v$, $A=a$. Lemma \ref{l:linq} says
that the right hand sides of both lattices are linear in any
variable. Several equations for the coef\/f\/icients follow from
the relations for the functions $b$, $c$:
\begin{gather*}
 b(u,u_1)= a_{11}uu_1+a_{01}u_1+f_1(u)= \eps a_{11}uu_1+\eps a_{01}u-p_1(u_1),
 \\
 c(v,v_1)= \eps a_{11}vv_1+\eps a_{10}v-g_1(v_1)= a_{11}vv_1+a_{10}v_1+q_1(v).
\end{gather*}
After the f\/inal check of the compatibility condition we arrive
to the pair (\ref{X4},\ref{Y4}).
\end{proof}

In order to f\/inish the classif\/ication we have to consider the
lattices with the function $a$ of the form $a(u,v)=a_1(u)a_2(v)$.

\begin{proposition}\label{pr:a0}
The consistent lattices with the condition $aa_{uv}=a_ua_v$, are
brought by the admissible substitutions to one of the pairs
$(\ref{X5},\ref{Y5})$, $(\ref{X6},\ref{Y6})$ or
$(\ref{X8},\ref{Y8})$.
\end{proposition}
\begin{proof}
The substitution (\ref{newuv}) allows to reduce the problem
(temporarily) to the case $a=1$. Then the compatibility conditions
for the systems (\ref{F})--(\ref{Q}) are
\begin{gather}\label{a0:hrbc}
 \frac{h_v}{h}=b_{u_1}+b\frac{h_{u_1}}{h},\qquad
 \eps\frac{h_{u_1}}{h}=c_v+c\frac{h_v}{h},\qquad
 \eps\frac{r_{v_1}}{r}=b_u+b\frac{r_u}{r},\qquad
 \frac{r_u}{r}=c_{v_1}+c\frac{r_{v_1}}{r}.
\end{gather}
As a corollary, the equations hold
\begin{gather*}
 b_u(\log h)_{vu_1}=0,\qquad c_{v_1}(\log h)_{vu_1}=0,\qquad
 b_{u_1}(\log r)_{uv_1}=0,\qquad c_v(\log r)_{uv_1}=0.
\end{gather*}
We will use also the following consequences of the relations
(\ref{ahPQ}), (\ref{arFG}):
\begin{gather}\label{a0:k12}
 P_u+Q_v+k_1=0,\qquad F_u+G_v+k_2=0.
\end{gather}

1) At f\/irst, assume that both quantities $(\log h)_{vu_1}$ and
$(\log r)_{uv_1}$ do not vanish. Then $b$, $c$ are constants and
$bc=\eps$. A scaling of $u$ and $v$ allows to set $b=-1$,
$c=-\eps$. One obtains, after the integration of the equations
(\ref{F})--(\ref{Q}) and taking (\ref{a0:k12}) into account, the
lattices of the form
\begin{gather*}
 u_x =f(v-u_1)+\alpha_1u+\alpha,\qquad
 v_x =\eps f(v_{-1}-u)+\beta_1v,
 \\
 u_y =p(\eps u_{-1}-v)+\gamma_1u+\gamma, \qquad
 v_y =p(\eps u-v_1)+\delta_1v.
\end{gather*}
The direct computation proves (notice that $f''p''\ne0$ by
assumption) that the linear terms are zero and brings to the pair
(\ref{X8},\ref{Y8}).

2) Now let either $(\log h)_{vu_1}=0$ or $(\log r)_{uv_1}=0$.
Taking the symmetry (\ref{sym_xy}) into account, we assume for
def\/initeness that $h=m(u_1)n(v)$. Then the two f\/irst equations
(\ref{a0:hrbc}) take the form
\begin{gather*}
 \frac{n'}{n}=b_{u_1}+b\frac{m'}{m},\qquad \eps\frac{m'}{m}=c_v+c\frac{n'}{n},
\end{gather*}
whence
\begin{gather*}
 m'=\mu m,\qquad n'=\nu n,\qquad b_{u_1}=\nu-\mu b,\qquad c_v=\eps\mu-\nu c.
\end{gather*}
Here we have to consider several subcases.

2.1) Let $\mu\nu\ne0$. After scaling, assume $\mu=1$, $\nu=-1$,
$h=e^{u_1-v}$. Then
\begin{gather*}
 b=e^{-u_1}\bar m(u)-1,\qquad c=e^v\bar n(v_1)-\eps,
\end{gather*}
and two last equations (\ref{a0:hrbc}) are reduced to relations
\begin{gather*}
 \bar m'(u)r+\bar m(u)r_u=0,\qquad \bar n'(v_1)r+\bar n(v_1)r_{v_1}=0,\qquad
 r_u+\eps r_{v_1}=0.
\end{gather*}
If $\bar m=\bar n=0$ then we obtain the lattice of the form
(\ref{X8},\ref{Y8}). Otherwise,
\begin{gather*}
 b=\beta e^{-\delta\eps u-u_1}-1,\qquad c=\gamma e^{v+\delta v_1}-\eps,\qquad
 r=\lambda e^{\delta(\eps u-v_1)},
\end{gather*}
where at least one of the coef\/f\/icients $\beta$, $\gamma$ is not
zero. The solution of equations (\ref{F}), (\ref{G}) is
\begin{gather*}
 F=e^{u_1-v}-\beta e^{-\delta\eps u-v}+f(u),\qquad
 G=\eps e^{u-v_{-1}}-\gamma e^{u+\delta v}+g(v).
\end{gather*}
The functions $P$, $Q$ are easily found as well. One easily
proves, by use of (\ref{a0:k12}), that $\delta=-1$, $\eps=1$,
$\beta=\gamma$ and the functions $f,g$ are linear. After the
direct computation of the constants and the substitution $e^u\to
u$, $e^{-v}\to v$ (of course, it spoils the gauge $a=1$) we come
to the pair (\ref{X5},\ref{Y5}).

2.2) Let $\mu=0$, $\nu\ne0$. Taking $\nu=1$, $h=e^v$ we obtain
\begin{gather*}
 b=u_1+\bar m(u),\qquad c=e^{-v}\bar n(v_1),
 \\[.5ex]
 r_u=0,\qquad \bar n'(v_1)r+\bar n(v_1)r_{v_1}=0,\qquad
 \eps r_{v_1}=\bar m'(u)r.
\end{gather*}
From here, $b=u_1+\eps\delta u+\beta$, $c=\gamma e^{-v-\delta
v_1}$, $r=\lambda e^{\delta v_1}$ and the solution of equations
(\ref{F}), (\ref{G}) is
\begin{gather*}
 F=(u_1+\delta\eps u+\beta)e^v+f(u),\qquad
 G=-\eps e^{v_{-1}}-\gamma ue^{-\delta v}+g(v).
\end{gather*}
The second equation (\ref{a0:k12}) now reads $\delta\eps
e^v+f'(u)+\gamma\delta ue^{-\delta v}+g'(v)+k_2=0$. Since
$\gamma\ne0$ in virtue of the nondegeneracy condition (\ref{ne0}),
hence $\delta=0$. Taking $r=1$ and solving equations (\ref{P}),
(\ref{Q}) one obtains
\begin{gather*}
  P=-\eps u_{-1}-(u+\beta)v+p(u),\qquad Q=v_1+\gamma ue^{-v}+q(v).
\end{gather*}
However, then the f\/irst equation (\ref{a0:k12}) implies
$\gamma=0$. Therefore, this case is empty (as well as the case
$\mu\ne0$, $\nu=0$, due to the symmetry (\ref{sym_uv})).

2.3) Finally, let $\mu=\nu=0$. Then $b=b(u)$, $c=c(v_1)$ and one
can take $h=1$. Then $F=u_1+b(u)v+f(u)$, $G=-\eps
v_{-1}-c(v)u-g(v)$ and substitution into (\ref{a0:k12}) yields
\begin{gather*}
 b'(u)v+f'(u)-c'(v)u-g'(v)+k_2=0.
\end{gather*}
From here,
\begin{gather*}
 b(u)=\alpha u^2+2\beta_1u+\beta,\qquad c(v)=\alpha v^2+2\gamma_1v+\gamma.
\end{gather*}
Moreover, elimination of $r$ from the equations (\ref{a0:hrbc})
brings to the relation
\begin{gather*}
 \eps b''+(b'^2-bb'')c=\eps^2c''+\eps(c'^2-cc'')b,\qquad
 b=b(u),\qquad c=c(v_1).
\end{gather*}
Denote $\Delta_b=b'^2-2bb''=4(\beta^2_1-\alpha\beta)$, and
analogously for $c$, then this relation takes the form
\begin{gather*}
 2\eps\alpha+(\Delta_b+2\alpha b)c=2\eps^2\alpha+\eps(\Delta_c+2\alpha c)b.
\end{gather*}
If $\alpha\ne0$ then it follows that $\eps=1$ and
$\Delta_b=\Delta_c=0$, that is $b$ and $c$ are full squares. The
linear change of variables brings them to the form $b=u^2$,
$c=v^2_1$. Then one f\/inds from (\ref{a0:hrbc}) that
$r=(uv_1-1)^{-2}$ and solution of the equations (\ref{P}),
(\ref{Q}) is $P=u_{-1}/(u_{-1}v-1)+p(u)$, $Q=-v_1/(uv_1-1)+q(v)$.
Equations (\ref{a0:k12}) say that functions $f,g,p,q$ are linear
and after the direct computation of the constants we obtain the
pair (\ref{X6},\ref{Y6}).

If $\alpha=0$ then $\beta_1\gamma_1=0$,
$\beta^2_1\gamma=\eps\gamma^2_1\beta$. Since functions $b,c$ do
not vanish in virtue of (\ref{ne0}), hence $\beta_1=\gamma_1=0$,
that is the lattice (\ref{uvx}) is linear. Direct calculations
prove that either the second lattice is linear as well, or we
arrive to a particular case of the pair (\ref{X8},\ref{Y8}).
\end{proof}

\section{Associated equations}\label{sec:assoc}

\subsection{Hyperbolic PDE systems}\label{sec:hyps}

In accordance to the nondegeneracy conditions the equations
(\ref{uvx}), (\ref{uvy}) can be solved with respect to the
variables $u_{\pm1}$, $v_{\pm1}$:
\begin{gather*}
 u_1=\tilde F(u_x,u,v),\qquad  v_{-1}=\tilde G(u,v,v_x),
 \\
 u_{-1}=\tilde P(u_y,u,v),\qquad  v_1 =\tilde Q(u,v,v_y).
\end{gather*}
Therefore, these variables can be eliminated from the expressions
for the mixed derivatives and some system of partial
dif\/ferential equations appears. The original lattices def\/ine
its B\"acklund auto-transformation. The form of the system is
given by equations
\begin{gather*}
 u_{xy}=f_4u_xu_y+f_3u_x+f_2u_y+f_1,\qquad
 v_{xy}=g_4v_xv_y+g_3v_x+g_2v_y+g_1,
\end{gather*}
where the coef\/f\/icients depend on $u$, $v$. Indeed, consider the
equalities
\begin{gather*}
 u_{xy}=F_{u_1}T(P)+F_uu_y+F_vv_y=P_{u_{-1}}T^{-1}(F)+P_uu_x+P_vv_x.
\end{gather*}
In the f\/irst one the variables $u_1$, $v_1$ should be eliminated
only and we see that the expression for $u_{xy}$ does not depend
on $v_x$ and is linear in $u_y$. Analogously, elimination of
$u_{-1}$, $v_{-1}$ in the second equality proves that the
expression for $u_{xy}$ does not depend on $v_y$ and is linear in
$u_x$. The formula for $v_{xy}$ is proved analogously.

\begin{proposition}\label{pr:hyps}
In virtue of the lattices from the list {\rm \ref{list}}, the
variables $u$, $v$ satisfy the respective systems from the list
{\rm \ref{list:hyps}}.
\end{proposition}

\subsection[Ruijsenaars-Toda lattices]{Ruijsenaars--Toda lattices}\label{sec:RTL}

As we have already explained in the Introduction, elimination of
the variable $v$ allows to rewrite the lattice (\ref{uvx}) in the
form of Ruijsenaars--Toda type lattice (\ref{RTLxx}), and its
symmetry (\ref{uvy}) in the form (\ref{RTLxy}). Moreover, the
equation (\ref{RTLyy}) is fulf\/illed as well. The triples of the
lattices corresponding to the list \ref{list} are enumerated in
the list \ref{list:RTL}. The remarkable property is that
elimination of $u$ instead of $v$ brings to the same equations.
This can be proved without calculations since it is easy to see
that the lattices from the list \ref{list} are invariant with
respect to the involution
\begin{gather*}
 u_n\leftrightarrow\sigma v_{-n},\qquad x\to-x,\qquad y\to-y,
\end{gather*}
and the lattices from the list \ref{list:RTL} are invariant with
respect to the involution
\begin{gather*}
 u_n\leftrightarrow\sigma u_{-n},\qquad x\to-x,\qquad y\to-y,
\end{gather*}
where $\sigma=1$ for (\ref{X1}), (\ref{Y1}), (\ref{R1}) and
$\sigma=-1$ in other cases. It should be noted that this property
is not invariant under the general substitutions (\ref{newuv}).
For an arbitrary choice of the variables $u$ and $v$ the
corresponding Ruijsenaars--Toda lattices coincide after a suitable
point substitution. As to the lattices (\ref{X7}), (\ref{Y7}),
(\ref{X8}), (\ref{Y8}), they possess the above symmetry only if
$\eps=1$.

\subsection{Discrete Toda lattices}\label{sec:DTL}

Coincidence of the lattices (\ref{RTLxx}) for the variables $u$
and $v$ allows to expand the equations on the square grid and
leads to the discrete Toda lattices. More precisely, this property
means that the equations (\ref{uvx}), (\ref{uvy}) def\/ine the
B\"acklund auto-transformations for the lattices (\ref{RTLxx}),
(\ref{RTLyy}) correspondingly. This allows to introduce the
variable $u(n,m,x,y)$, so that the pair $(u(n),v(n))$ is
identif\/ied with $(u(n,m),u(n,m+1))$ at arbitrary $m$. Let us
rewrite the lattices from the list~\ref{list} denoting the shifts
by the double subscripts:
\begin{gather}
\label{umnx}
 u_x=F(u_{0,1},u,u_{1,0})=G(u_{-1,0},u,u_{0,-1}),
 \\[.5ex]
\label{umny}
 u_y=P(u_{0,-1},u,u_{1,0})=Q(u_{-1,0},u,u_{0,1}).
\end{gather}
Then the following properties are valid:

1) equations $F=G$ and $P=Q$ are equivalent and can be rewritten
as the discrete Toda lattice
\begin{gather}\label{DTL}
 f(u_{0,1},u)+\tilde f(u_{0,-1},u)=g(u_{1,0},u)+\tilde g(u_{-1,0},u);
\end{gather}

2) equation (\ref{DTL}) is consistent with the dynamics on $x$ and
$y$, that is, the equations obtained by dif\/ferentiating of this
equation in virtue of (\ref{umnx}) or (\ref{umny}) are its
consequences;

3) the variables $u(n)=u(m,n)$ for any $m$ satisfy the
Ruijsenaars--Toda lattices (\ref{RTLxx}), (\ref{RTLyy});

4) the variables $u(m)=u(m,n)$ for any $n$ satisfy the analogous
lattices (possibly, with the dif\/ferent right hand sides);

5) the variables along the diagonals $u(n)=u(m-n,n)$ satisfy a
Toda type lattice with respect to the $x$-derivatives
\begin{gather*}
 u_{xx}=\tilde A(u_x,u_1,u,u_{-1}),
\end{gather*}
and the variables along the complementary diagonals
$u(n)=u(m+n,n)$ satisfy a Toda type lattice with respect to the
$y$-derivatives
\begin{gather*}
 u_{yy}=\tilde C(u_y,u_1,u,u_{-1}).
\end{gather*}

6) all these equations are Lagrangian.

\section{Non-Hamiltonian lattices}\label{sec:eps}

The pairs (\ref{X7},\ref{Y7}) and (\ref{X8},\ref{Y8}) are of
interest as counter-examples to the statement that existence of
one symmetry implies the existence of the whole hierarchy. These
lattices do not possess the higher symmetries at arbitrary $f$,
$g$ and $\eps$. In particular, the necessary integrability
conditions fail for the associated lattices of the form
(\ref{RTLxx}). In accordance to \cite{Yamilov2000} the simplest of
these conditions reads
\begin{gather*}
 D_x(\log A_{u_{1,x}})\in\mathop{\rm Im}(T-1).
\end{gather*}
The direct computation proves that elimination of $v$ from
equations (\ref{X7}) brings to the lattice
\begin{gather*}
 u_{xx}=\frac{(u_x+1)(\eps u_x-1)}{1+\eps}
  \left(\frac{u_{1,x}+1}{u_1+\eps u}
  -\frac{\eps u_{-1,x}-1}{u+\eps u_{-1}}\right)
\end{gather*}
and the above condition is fulf\/illed for this lattice only if
$\eps=1$.

It turns out, however, that the nature of these counter-examples
is not too profound. It is easy to see that the non-invertible
substitution $U=v-u_1$, $V=\eps u-v_1$ reduces the lattices
(\ref{X8},\ref{Y8}) to the disjoint pairs
\begin{gather*}
 U_x=\eps f(U_{-1})-f(U_1),\qquad U_y=0;\qquad\qquad
 V_x=0,\qquad V_y=\eps p(V_{-1})-p(V_1).
\end{gather*}
Analogous, but not so evident substitution exists for the pair
(\ref{X7},\ref{Y7}) as well: the variables
\begin{gather}\label{eps:UV}
 U=\frac{(1+\eps)(u+v)}{(\eps v_{-1}-u)(\eps v-u_1)},\qquad
 V=\frac{(1+\eps)(u+v)}{(\eps u_{-1}-v)(\eps u-v_1)}
\end{gather}
satisfy the equations
\begin{gather}\label{eps:UxVy}
 U_x=U(U_1+U-\eps(U+U_{-1})),\qquad U_y=0;\nonumber
 \\
 V_x=0,\qquad V_y=-V(V_1+V-\eps(V+V_{-1})).
\end{gather}
In both cases the quantities $U(n)$ are the local f\/irst
integrals for one lattice in the pair, and~$V(n)$ for another one.
In these variables the consistency of $x$- and $y$-f\/lows become
trivial and is not related to integrability (although at $\eps=1$
and for some $f$, $p$ equations may ``accidentally'' coincide with
integrable Volterra type lattices \cite{Yamilov1983}).

The fact that the local f\/irst integrals satisfy the closed
lattice with respect to the second independent variable can be
easily explained in the framework of the associated hyperbolic
system. For the pair (\ref{X7},\ref{Y7}) it is of the form
\begin{gather}\label{eps:xy}
 u_{xy}=\frac{(u_x+1)(u_y+1)}{u+v},\qquad v_{xy}=\frac{(v_x-1)(v_y-1)}{u+v}
\end{gather}
(notice that the units may be dropped, due to the substitution
$u\to u-x-y$, $v\to v+x+y$, however this leads to nonautonomous
lattices). This is the Liouville equation type system
\cite{Zhiber_Ibragimov_Shabat_1979, Sokolov_Zhiber_1995,
Zhiber_Sokolov_2001}. Elimination of the shifts in the f\/irst
integrals (\ref{eps:UV}) leads to the invariants of this system,
that is, to the quantities $I(u,v,u_x,v_x,\dots)$,
$J(u,v,u_y,v_y,\dots)$ which satisfy the properties $I_y=0$,
$J_x=0$ in virtue of (\ref{eps:xy}). It can be proved that, in the
case of the hyperbolic system of the second order, all solutions
of the equation $I_y=0$ are expressed through the $x$-derivatives
of at most two basic invariants. Therefore, the invariants $U_1$,
$U$, $U_{-1}$ must be related by some dif\/ferential constraint,
as the f\/irst equation (\ref{eps:UxVy}) shows.

Finally, notice that the system (\ref{eps:xy}) implies
\begin{gather*}
 (\log u_{xy})_{xy}=0,\qquad (\log v_{xy})_{xy}=0.
\end{gather*}
This gives the formula for the general solution
\begin{gather*}
 u=a(x)b(y)+c(x)+d(y),\qquad v=-u+\frac{(u_x+1)(u_y+1)}{u_{xy}}
\end{gather*}
under assumption $a'b'\ne0$, and this ansatz reduces the lattices
(\ref{X7},\ref{Y7}) to the recurrent relations
\begin{gather*}
 a_1=\frac{\eps c'-1}{a'},\qquad c_1=aa_1-\eps c;\qquad
 b_1=\frac{d'+1}{b'},\qquad d_1=\eps(bb_1-d).
\end{gather*}
These relations are equivalent to some Volterra type lattice, for
example the substitution $a\to\eps^na$ and scaling of $x$ brings
to the equation (known to be integrable at $\eps=1$
\cite{Yamilov1983})
\begin{gather*}
 \eps^na_{n,x}=\frac{1}{a_{n+1}-a_{n-1}}.
\end{gather*}

\section{Concluding remarks}\label{sec:n}

The integrable lattices of the form (\ref{uvx}) or (\ref{RTLxx})
admit the very important generalization: their right hand sides
may contain parameters which depend on $n$ and def\/ine the
discrete spectra added by the iterated B\"acklund transformations.
Such generalizations are known, actually, for all lattices from
the above lists, see e.g. \cite{AdlerYamilov1994,
AdlerShabat1998}. It turns out, however, that $n$-dependent
lattices (\ref{uvx}) {\it are not} consistent with the symmetries
of the form (\ref{uvy}), only with the higher symmetries of NLS
type. The nature of this phenomena is not well understood for the
present. It takes place in another situations as well, for
example, the dressing chain
$v_{n,x}+v_{n+1,x}=(v_n-v_{n+1})^2+\alpha_n$ is consistent with
equation $v_{n,y}=(v_{n+1}-v_{n-1})^{-1}$ only if
$\alpha_n=\alpha_{n+1}$ \cite{AdlerShabat2006}. However, the
exceptional pairs (\ref{X7},\ref{Y7}), (\ref{X8},\ref{Y8}) may
depend on $n$, for example, the following pair is consistent:
\begin{gather*}
  u_{n,x}=\frac{v_n+(1+\eps_n)u_n+u_{n+1}}{\eps_nv_n-u_{n+1}},  \qquad
  v_{n,x}=\frac{v_{n-1}+(1+\eps^{-1}_{n-1})v_n+u_n}{v_{n-1}-\eps^{-1}_{n-1}u_n},
               \\
  u_{n,y}=\frac{v_n+(1+\eps^{-1}_{n-1})u_n+u_{n-1}}{\eps^{-1}_{n-1}v_n-u_{n-1}},   \qquad
  v_{n,y}=\frac{v_{n+1}+(1+\eps_n)v_n+u_n}{v_{n+1}-\eps_nu_n}.
\end{gather*}

Another sort of the nonautonomous generalizations is related to
the master-symmetries of the lattices (\ref{uvx}), (\ref{uvy}),
see \cite{AdlerShabatYamilov2000, Nijhoff_Hone_Joshi}.

The multif\/ield lattices also should be mentioned. Probably, the
earliest example of such kind was found in
\cite{Svinolupov_Yamilov_1991, MRZ}. It generalizes the pair
(\ref{X6},\ref{Y6}) in the rational form:
\begin{gather*}
  u_x=u_1+\langle u,v\rangle u,\qquad
 -v_x=v_{-1}+\langle u,v\rangle v; \qquad
  u_y=\frac{u_{-1}}{\langle u_{-1},v\rangle -1},\qquad
 -v_y=\frac{v_1}{\langle u,v_1\rangle -1}.
\end{gather*}
The vectors $u,v$ satisfy, in virtue of these equations, the
system
\begin{gather*}
 u_{xy}= \langle u,v\rangle u_y+(\langle u_y,v\rangle -1)u,\qquad
 v_{xy}=-\langle u,v\rangle v_y-(\langle u,v_y\rangle +1)v.
\end{gather*}
However, an analog of the lattice (\ref{R6}) is absent in this
example.

\section{Lists}\label{sec:lists}

\subsection{Consistent pairs of the Hamiltonian lattices}\label{list}
\vspace{-4ex}
\begin{gather*}
\ltag{X1}{X_1}
 u_x=\frac{a}{v-u_1}-\frac{a_v}{2}, \qquad v_x=\frac{a}{v_{-1}-u}+\frac{a_u}{2}, \\
\ltag{Y1}{Y_1}
u_y=\frac{a}{u_{-1}-v}+\frac{a_v}{2}, \qquad v_y=\frac{a}{u-v_1}-\frac{a_u}{2}, \\
a=\alpha_1u^2v^2+\alpha_2uv(u+v)+\alpha_3(u^2+v^2)+\alpha_4uv+\alpha_5(u+v)+\alpha_6,
\\
H=\log(u_1-v)-\frac12\log a, \qquad R=-\log(u-v_1)+\frac12\log a,
\\[1.5ex]
\ltag{X2}{X_2}
u_x=(u-u_1)(u-v), \qquad v_x=(v_{-1}-v)(u-v), \\
\ltag{Y2}{Y_2} u_y=\frac{u-u_{-1}}{v-u_{-1}}, \qquad
v_y=\frac{v_1-v}{v_1-u} ,
\\
a=u-v,\qquad H=uv-u_1v,\qquad R=\log(u-v_1)-\log(u-v),
\\[1.5ex]
\ltag{X3}{X_3} u_x=(1+e^{u_1-u})(1+e^{u-v}),    \qquad
v_x=(1+e^{v-v_{-1}})(1+e^{u-v}),
\\
\ltag{Y3}{Y_3} u_y=\dfrac{1+e^{u_{-1}-u}}{1-e^{u_{-1}-v}},
\qquad v_y=\dfrac{1+e^{v-v_1}}{1-e^{u-v_1}},
\\
a=1+e^{v-u},\qquad H=-e^{u_1-v}-e^{u-v},\qquad
R=\log(1-e^{v_1-u})-\log(1+e^{v-u}),
\\[1.5ex]
\ltag{X4}{X_4} u_x=e^{u_1-u}+e^{u_1-v},\qquad
v_x=e^{v-v_{-1}}+e^{u-v_{-1}},
\\
\ltag{Y4}{Y_4} u_y=e^{u_{-1}-u}+e^{u_{-1}-v}, \qquad
v_y=e^{v-v_1}+e^{u-v_1},
\\
a=1+e^{v-u},\qquad H=-e^{u_1-v},\qquad R=-e^{u-v_1},
\\[1.5ex]
\ltag{X5}{X_5} u_x=e^{u_1-v}+e^{u-v}, \qquad
v_x=e^{u-v_{-1}}+e^{u-v},
\\
\ltag{Y5}{Y_5} u_y=e^{v-u_{-1}}+e^{v-u},  \qquad
v_y=e^{v_1-u}+e^{v-u} ,
\\
a=1,\qquad H=-e^{u_1-v}-e^{u-v},\qquad R=e^{v_1-u}+e^{v-u},
\\[1.5ex]
\ltag{X6}{X_6} u_x=e^{u_1-u}+e^{u-v},   \qquad
v_x=e^{v-v_{-1}}+e^{u-v},
\\
\ltag{Y6}{Y_6} u_y=\frac{e^{u_{-1}-u}}{e^{u_{-1}-v}-1},\qquad
v_y=\frac{e^{v-v_1}}{e^{u-v_1}-1},
\\
a=-e^{v-u},\qquad H=e^{u_1-v}+\frac12e^{2(u-v)},\qquad
R=\log(1-e^{u-v_1}).
\end{gather*}

\subsection{Hyperbolic PDE systems}\label{list:hyps}
\vspace{-4ex}
\begin{gather*}
au_{xy}=a_uu_xu_y+2\tilde a(u_x-u_y)+\tilde aa_v-\tilde a_va,
\\
av_{xy}=a_vv_xv_y-2\tilde a(v_x-v_y)+\tilde aa_u-\tilde
a_ua,\ltag{H1}{H_1}
\\
4\tilde a=aa_{uv}-a_ua_v,
\end{gather*}
\begin{alignat}{2}
\ltag{H2}{H_2}
 & u_{xy}=\frac{(u_y-1)u_x}{u-v}+u_y(u-v), &&
   v_{xy}=-\frac{(v_y-1)v_x}{u-v}-v_y(u-v), \\
\ltag{H3}{H_3}
 & u_{xy}=-\frac{u_x(u_y-1)}{1+e^{u-v}}+u_y(1+e^{u-v}), &\qquad &
   v_{xy}= \frac{v_x(v_y-1)}{1+e^{u-v}}-v_y(1+e^{u-v}), \\
\ltag{H4}{H_4}
 & u_{xy}=-\frac{u_xu_y}{1+e^{u-v}}+1+e^{u-v},&&
   v_{xy}= \frac{v_xv_y}{1+e^{u-v}}-1-e^{u-v},\\
\ltag{H5}{H_5}
 & u_{xy}=e^{u-v}u_y-e^{v-u}u_x, && v_{xy}=e^{v-u}v_x-e^{u-v}v_y, \\
\ltag{H6}{H_6}
 & u_{xy}=-u_xu_y+2e^{u-v}u_y-1, && v_{xy}=v_xv_y-2e^{u-v}v_y+1.
\end{alignat}

\subsection[Ruijsenaars-Toda lattices]{Ruijsenaars--Toda lattices}\label{list:RTL}
\vspace{-2ex}
\begin{align}
\ltag{R1}{R_1}
 &\left\{\begin{aligned}
 & u_{xx}= (u_x^2+r(u))\left(\frac{u_{1,x}}{T(a)}-\frac{u_{-1,x}}{a}
  +\frac{1}{2}\partial_u\log(T(a)a)\right)- \frac{1}{2}r'(u), \\
 & u_{yy}= (u_y^2+r(u))\left(\frac{u_{1,y}}{T(a)}-\frac{u_{-1,y}}{a}
  +\frac{1}{2}\partial_u\log(T(a)a)\right)- \frac{1}{2}r'(u), \\
 & (u_xu_y-r(u))(u_1-u_{-1})+(u_x+u_y)(a+\frac12a_{u_{-1}}(u_1-u_{-1}))=0 \\
 & a=a(u,u_{-1}),\qquad 4r(u)=2aa_{u_{-1}u_{-1}}-a^2_{u_{-1}},
\end{aligned}\right.\\[0mm]
\ltag{R2}{R_2}
 &\left\{\begin{aligned}
 & u_{xx}=u_x\left(\frac{u_{1,x}}{u_1-u}-\frac{u_{-1,x}}{u-u_{-1}}
    -u_1+2u-u_{-1}\right),\\
 & u_{yy}=-u_y(u_y-1)\left(\frac{u_{1,y}}{u_1-u}-\frac{u_{-1,y}}{u-u_{-1}}\right),\\
 & u_xu_y=(1-u_y)(u_1-u)(u-u_{-1}),
\end{aligned}\right.\\[0mm]
\ltag{R3}{R_3}
 &\left\{\begin{aligned}
 & u_{xx}=u_x\left(\frac{u_{1,x}}{1+e^{u-u_1}}
    -\frac{u_{-1,x}}{1+e^{u_{-1}-u}}-e^{u_1-u}+e^{u-u_{-1}}\right), \\
 & u_{yy}=-u_y(u_y-1)\left(\frac{u_{1,y}}{1+e^{u-u_1}}
    -\frac{u_{-1,y}}{1+e^{u_{-1}-u}}\right), \\
 & u_xu_y=(u_y-1)(e^{u_1-u}+1)(e^{u-u_{-1}}+1),
\end{aligned}\right.\\[0mm]
\ltag{R4}{R_4}
 &\left\{\begin{aligned}
 & u_{xx}=u_x(u_{1,x}-u_{-1,x}-e^{u_1-u}+e^{u-u_{-1}}), \\
 & u_{yy}=u_y(u_{-1,y}-u_{1,y}-e^{u_{-1}-u}+e^{u-u_1}), \\
 & u_x=u_ye^{u_1-u_{-1}},
\end{aligned}\right.\\[0mm]
\ltag{R5}{R_5}
 &\left\{\begin{aligned}
 & u_{xx}=\frac{u_{1,x}u_x}{1+e^{u-u_1}}-\frac{u_xu_{-1,x}}{1+e^{u_{-1}-u}}, \\
 & u_{yy}=\frac{u_{1,y}u_y}{1+e^{u-u_1}}-\frac{u_yu_{-1,y}}{1+e^{u_{-1}-u}}, \\
 & u_xu_y=(e^{u_1-u}+1)(e^{u-u{-1}}+1),
\end{aligned}\right.\\[0mm]
\ltag{R6}{R_6}
 &\left\{\begin{aligned}
 & u_{xx}=e^{u_1-u}u_{1,x}-e^{u-u_{-1}}u_{-1,x}-e^{2(u_1-u)}+e^{2(u-u_{-1})},\\
 & u_{yy}=u^2_y(e^{u_1-u}u_{1,y}-e^{u-u_{-1}}u_{-1,y}), \\
 & u_x-1/u_y=e^{u_1-u}+e^{u-u{-1}}.
\end{aligned}\right.
\end{align}

\subsection*{Acknowledgements}

This research was supported by the RFBR grant \# 04-01-00403.

\LastPageEnding


\begin{thebibliography}{99}
\footnotesize

\bibitem{Ruijsenaars1986} Ruijsenaars S.N.M., Relativistic Toda system, Preprint
Stichting Centre for Mathematics and Computer Sciences, Amsterdam,
1986.

\bibitem{Ruijsenaars1990} Ruijsenaars S.N.M., Relativistic Toda systems,
{\it Comm. Math. Phys.}, 1990, V.133, 217--247.

\bibitem{BruschiRagnisco1989} Bruschi M., Ragnisco O.,
 On a new integrable Hamiltonian system with nearest-neighbour interaction,
 {\it Inverse Problems}, 1989, V.5, 983--998.

\bibitem{Suris1990} Suris Yu.B., Discrete time generalized Toda lattices:
 complete integrability and relation with relativistic Toda lattices,
 {\it Phys. Lett. A}, 1990, V.145, 113--119.

\bibitem{AdlerShabat1997a} Adler V.E., Shabat A.B.,
 On the one class of the Toda chains,
 {\it Theor. Math. Phys.}, 1997, V.111, 323--334.

\bibitem{AdlerShabat1997b} Adler V.E., Shabat A.B.,
 Generalized Legendre transformations,
 {\it Theor. Math. Phys.}, 1997, V.112, 935--948.

\bibitem{AdlerShabat1998} Adler V.E., Shabat A.B.,
 First integrals of the generalized Toda lattices,
 {\it Theor. Math. Phys.}, 1998, V.115, 349--358.

\bibitem{MarikhinShabat1999} Marikhin V.G., Shabat A.B., Integrable lattices,
{\it Theor. Math. Phys.}, 1999, V.118, 217--228.

\bibitem{AdlerMarikhinShabat2001} Adler V.E., Marikhin V.G., Shabat A.B.,
 Canonical B\"acklund transformations and Lagrangian chains,
{\it Theor. Math. Phys.}, 2001, V.129, 163--183.

\bibitem{RagniscoSantini1990} Ragnisco O., Santini P.M.,
A unif\/ied algebraic approach to integral and discrete evolution
equations, {\it Inverse Problems}, 1990, V.6, 441--452.

\bibitem{ShabatYamilov1991} Shabat A.B., Yamilov R.I.,
Symmetries of nonlinear lattices, {\it Leningrad Math. J.}, 1991,
V.2, 377--400.

\bibitem{AdlerYamilov1994} Adler V.E., Yamilov R.I.,
 Explicit auto-transformations of integrable chains,
 {\it J. Phys. A: Math. Gen.}, 1994, V.27, 477--492.

\bibitem{Yamilov2000} Yamilov R.I.,
 Symmetry approach to the classif\/ication from the point of view of
 the dif\/ferential-dif\/ference equations. Theory of transformations,
Doctoral Thesis, Ufa, 2000.

\bibitem{AdlerShabatYamilov2000} Adler V.E., Shabat A.B., Yamilov R.I.,
Symmetry approach to the integrability problem, {\it Theor. Math.
Phys.}, 2000, V.125, 1603--1661.

\bibitem{Adler2000a} Adler V.E.,
 Discretizations of the Landau--Lifshitz equation,
 {\it Theor. Math. Phys.}, 2000, V.124, 897--908.

\bibitem{Hirota} Hirota R., Nonlinear partial dif\/ference equations. II.
Discrete-time Toda equation, {\it J. Phys. Soc. Japan}, 1977,
V.43, 2074--2078.

\bibitem{Suris95} Suris Yu.B., Bi-Hamiltonian structure of the $qd$ algorithm
and new discretizations of the Toda lattice, {\it Phys. Lett. A},
1995, V.206, 153--161.

\bibitem{Adler2000b} Adler V.E., On the structure of the B\"acklund
transformations for the relativistic lattices, {\it J. Nonlinear
Math. Phys.}, 2000, V.7, 34--56,
\href{http://arxiv.org/abs/nlin.SI/0001072}{nlin.SI/0001072}.

\bibitem{AdlerSuris2004} Adler V.E., Suris Yu.B.,
 Q4: Integrable master equation related to an elliptic curve,
 {\it Internat. Math. Res. Not.}, 2004, V.47, 2523--2553, \href{http://arxiv.org/abs/nlin.SI/0309030}{nlin.SI/0309030}.

\bibitem{Zhiber_Ibragimov_Shabat_1979} Zhiber A.V., Ibragimov N.H., Shabat A.B.,
Liouville type equations, {\it DAN SSSR}, 1979, V.249, 26--29.

\bibitem{Sokolov_Zhiber_1995} Sokolov V.V., Zhiber A.V.,
On the Darboux integrable hyperbolic equations, {\it Phys. Lett.
A}, 1995, V.208, 303--308.

\bibitem{Zhiber_Sokolov_2001} Zhiber A.V., Sokolov V.V.,
Exactly solvable hyperbolic equations of the Liouville type, {\it
Uspekhi Mat. Nauk}, 2001, V.56, N~1, 63--106.

\bibitem{Yamilov1983} Yamilov R.I.,
 On classif\/ication of discrete evolution equations,
 {\it Uspekhi Mat. Nauk}, 1983, V.38, N~6, 155--156.

\bibitem{AdlerShabat2006} Adler V.E., Shabat A.B.,
 Dressing chain for the acoustic spectral problem,
 {\it Theor. Math. Phys.}, 2006, V.149, 1324--1337, \href{http://arxiv.org/abs/nlin.SI/0604008}{nlin.SI/0604008}.

\bibitem{Nijhoff_Hone_Joshi} Nijhof\/f F., Hone A., Joshi N.,
 On a Schwarzian PDE associated with the KdV hierarchy,
 {\it Phys. Lett. A}, 2000, V.267, 147--156, \href{http://arxiv.org/abs/solv-int/9909026}{solv-int/9909026}.

\bibitem{Svinolupov_Yamilov_1991} Svinolupov S.I., Yamilov R.I.,
The multi-f\/ield Schr\"odinger lattices, {\it Phys. Lett. A},
1991, V.160, 548--552.

\bibitem{MRZ} Merola I., Ragnisco O., Tu G.-Z.,
A novel hierarchy of integrable lattices, {\it Inverse Problems},
1994, V.10, 1315--1334,
\href{http://arxiv.org/abs/solv-int/9401005}{solv-int/9401005}.

\end{thebibliography}
\end{document}